\documentclass{aa}
\input{psfig}

\catcode`\@=11
\def\gsim{\ifmmode{\mathrel{\mathpalette\@versim>}}
    \else{$\mathrel{\mathpalette\@versim>}$}\fi}
\def\lsim{\ifmmode{\mathrel{\mathpalette\@versim<}}
    \else{$\mathrel{\mathpalette\@versim<}$}\fi}
\def\@versim#1#2{\lower 2.9truept \vbox{\baselineskip 0pt \lineskip
    0.5truept \ialign{$\m@th#1\hfil##\hfil$\crcr#2\crcr\sim\crcr}}}
\catcode`\@=12

\def\ei{{\it Einstein}}
\def\etal{\hbox{{\rm et al.}\,\,}}
\def\lx{$L_{\rm X}$}
\def\lb{$L_{\rm B}$}
\def\lsun{$L_{\odot}$}
\def\msun{$M_{\odot}$}
\def\parn{\par\noindent}
\def\sp{$\,\,$}

\def\ts{$kT_s$}
\def\th{$kT_h$}
\def\ref{\parn\hangindent=1truecm}

\begin{document}
\thesaurus{11.05.1, 11.09.1 NGC3923, 11.09.4, 13.25.2, 02.18.5}

\title{{\it BeppoSAX} observation of NGC3923, and the
problem of the X-ray emission in E/S0 galaxies of low and medium 
\lx/\lb}
\author{S. Pellegrini}
\offprints{S. Pellegrini}
\institute{
Dipartimento di Astronomia, Universit\`a di Bologna,
via Zamboni 33, I-40126 Bologna\\
 email: pellegrini@astbo3.bo.astro.it}
\date{Received...; accepted ...}

\titlerunning{{\it BeppoSAX} observation of NGC3923}
\maketitle
\begin{abstract}
We present the results of the analysis of the {\it BeppoSAX} LECS and MECS
pointed observation of the E4 galaxy NGC~3923, for which previous X-ray 
measurements had given  a medium X-ray to optical ratio \lx/\lb.
The spectral analysis over (0.5--10) keV reveals that the best 
representation of the data is the superposition of  
 two thermal components at temperatures of 0.4 keV and 6--8 keV.
The total emission is roughly equally divided between the two components,
over (0.5--4.5) keV.
Abundances are very subsolar at the best fit, but not constrained
by the data. The harder component is consistent with an origin 
from stellar sources; the softer component likely comes from hot gas. 
\lx\sp of this hot gas is not as large as expected for
a global inflow, in a galaxy of an optical luminosity as high as that 
of NGC~3923. So, it is suggested that a substantial amount of hot gas  was 
removed by internal agents, 
and that this process was helped by the flat mass distribution of the 
galaxy. Another possibility is that  gas was lost 
as a consequence of the episod of interaction or merger
that produced the system of shells visible in the optical band. 
Finally, the possible origins of the large scatter in the X-ray emission shown by galaxies of
similar \lb\sp are also reviewed. 
Like NGC~3923, many other low and medium \lx/\lb\sp galaxies reside in 
small groups, in which the ambient medium (if present) cannot strip them of
their hot gas; so, if only environmental factors are invoked to lower 
\lx/\lb, the most effective mechanism must be galaxy interactions.
The lower \lx/\lb\sp galaxies, though, are seen to occur across the whole 
range of galaxy densities.
Another possibility, to remove some or all of the hot gas, 
appeals to mechanisms internal to the galaxies, such as heating 
of the gas by supernovae explosions or accretion onto central massive 
black holes. This has been shown to work in general, for a large range of \lb, 
but it is not clear yet whether the hot gas abundances estimated from recent 
observations can be accomodated in it. 
\end{abstract}

\section{Introduction}
X-ray observations, beginning with the \ei\sp Observatory (Giacconi \etal 1979),
have demonstrated that normal early-type galaxies are X-ray emitters, with  0.2--4 keV
luminosities ranging from $\sim 10^{40}$ to $\sim 10^{43}$ erg s$^{-1}$
(Fabbiano 1989; Fabbiano et al. 1992). The X-ray luminosity
$L_{\rm X}$ is found to
correlate with the blue luminosity $L_{\rm B}$ ($L_{\rm X}\propto L_{\rm B}^
{2.0\pm 0.2}$),  but there is a large scatter of roughly two orders of magnitude
in \lx\sp at any fixed \lb$> 3\times 10^{10}$\lsun.
The analysis of the \ei\sp spectral data already revealed that 
in the X-ray brightest objects the X-ray radiation comes from thermal emission from a hot,
optically thin gas, at a temperature of $\sim 1$ keV (Canizares et al. 1987, hereafter CFT). It also revealed that the X-ray
emission temperature increases with decreasing \lx/\lb, until 
the dominant contribution to the total emission comes from a hard thermal component, 
similar to that dominating the emission in spiral galaxies (Kim et al. 1992). Since a population of low mass X-ray binaries 
(LMXB)
can explain the X-ray emission of the bulge of M31, and that of bulge-dominated 
spirals, it is likely that in early-type galaxies an increasing 
fraction of the X-ray emission comes from stellar sources
as \lx/\lb\sp decreases.
Better quality spectra of low and medium \lx/\lb\sp galaxies 
(i.e., belonging to Group 1 or Group 2 in the analysis of
Kim et al. 1992, or with \lx/\lb$<30.3$ in the {\it Einstein} band) have
been made available  recently for several galaxies, thanks to {\it ROSAT}
 and {\it ASCA} pointed observations. 

The analysis of {\it ROSAT} data confirmed the findings based on the 
{\it Einstein} data for several low and medium \lx/\lb\sp galaxies
(see, e.g., the sample of 61 early-type  galaxies
observed with the {\it ROSAT} PSPC built by Irwin \& Sarazin 1998). 
The emerging  picture was the existence of at least two spectral components, a
soft one likely due to hot gas or stellar sources, and a hard one, whose temperature is not well constrained,
due to the lack of sensitivity of the PSPC above 2 keV.

Sensitive over (0.5--10) keV, {\it ASCA}
pointed $\sim 30$ early-type galaxies, among which there
are a handful with low or medium $L_{\rm X}/L_{\rm B}$
(Matsushita et al. 1994; Kim et al. 1996; Matsumoto et al. 
1998; Buote \& Fabian 1998, hereafter BF). In all the 12 E/S0s of their sample
 Matsumoto et al. find a variable amount of soft thermal emission, of temperature ranging from 
0.3 to 1 keV, coupled to hard emission.
The amount of this hard emission
roughly scales as the optical luminosity of the galaxies, and
is consistent with that of bulge-dominated spirals, such as M31. 
Using {\it ASCA} data over the 
energy range 0.5--5 keV, also BF find a hard component 
of $kT> 2$ keV coupled with a soft component,
in the low or medium $L_{\rm X}/L_{\rm B}$ galaxies of their sample.

The precise knowledge of the contribution of stellar sources to the total
X-ray emission is important to better constrain
the properties of the hot gas (i.e., temperature, abundance and luminosity), 
or to identify peculiarities such
as the presence of a mini-AGN, another possible contributor to the hard
emission (e.g., Colbert \& Mushotzky 1998). So,  stronger constraints
on the gas flow phase and the galaxy properties can be derived.
The contribution of stellar sources can be best evaluated with observations over a large
bandwidth.
In this paper we report on the X-ray emission from the E4 galaxy NGC~3923,
as detected by the {\it BeppoSAX} satellite over (0.5--10) keV.
Our purpose is to investigate in detail the nature of the X-ray emission in
this medium \lx/\lb\sp galaxy ({\it Einstein} observations showed that
log$(L_{\rm X}/L_{\rm B})=30.24$ for it, Fabbiano \etal 1992).
The {\it BeppoSAX} observation of NGC~3923,
exploiting a moderate spatial resolution coupled with the
large energy band and the good spectral resolution of 
the satellite, permits the separation of the soft and the hard emission
components eventually present. 
The  spectral properties of NGC~3923 over 0.5--5 keV 
have been already studied by BF, using {\it ASCA} data referring to a radius 
of 2 arcmin (note that the radius encircling 80\% of the photons of a
point source is 3 arcmin for the {\it ASCA}-XRT). 
The {\it BeppoSAX} observation of NGC~3923 presents some advantages over that 
performed by {\it ASCA}:  a sharper PSF above 2 keV with respect to that of the 
{\it ASCA}-XRT, which is also asymmetrical\footnote{For example, the 
half power radius measured for the on-axis PSF at 6 keV is 
about a factor of two smaller than that of the {\it ASCA} on-axis PSF;
the comparison is even more
favorable at the 80\% and 90\% radii, due to the considerably reduced 
scattering of the MECS optics (Boella et al. 1997b).};
a lower instrumental background for the MECS with respect to the GIS,
which favors weak sources.
Finally, a larger covering in energy is presented here
with respect to that analysed by BF (i.e., 0.5--10 keV versus 0.5--5 keV),
 a spatial analysis is attempted over (1.7--10) keV, 
and a larger extraction region for the spectral data is used.
NGC~3923 has been pointed also by the {\it ROSAT} PSPC and HRI,
which allowed a detailed investigation of the spatial 
distribution of the X-ray emission (Buote \and Canizares 1998, hereafter 
BC).

This paper is organized as follows: we present in Sect. 2 the main properties
of NGC~3923, in Sect. 3 the results of the data analysis, in Sect. 4 the 
possible origins for the X-ray emissions of NGC~3923 are discussed in light
of the present results, and more in general 
 the possible origins of the large scatter in the X-ray emission shown 
by galaxies of similar \lb\sp are also reviewed;
 in Sect. 5 we summarize the  conclusions.

\newcommand{\tbsp}{\rule{0pt}{18pt}}
\begin{table*}
\caption[] { General characteristics of NGC~3923}
\begin{flushleft}
\begin{tabular}{ l  rl  l  l l  l  l l  l l  l }
\noalign{\smallskip}
\hline
\noalign{\smallskip}
Type$^a$ & RA     & Dec    & PA$^a$ & d$\, ^b$ &$B_{\rm T}^0 \, ^a$& log$L_{
\rm B} \, ^c$ & 
$R_{\rm e} \, ^a$         & $b/a \, ^a$ & $\sigma \, ^d$ & $N_{\rm H} \, ^e$ \\
         &(J2000) &(J2000) & (deg)  & (Mpc) &   (mag)    & ($L_{\odot}$) &
               &         & (km s$^{-1}$)         & (cm$^{-2}$)      \\
\noalign{\smallskip}
\hline
\noalign{\smallskip}
 E4 & $11^h 51^m 02^s.1$ & $-28^{\circ} 48^{\prime} 23^{\prime\prime} $ & 50 &
 41.5 & 10.27 & 11.3 & 49$\farcs$8 & 0.66 & 241 & 6.4$\times 10^{20}$ \\
\noalign{\smallskip}
\hline
\end{tabular} 
\end{flushleft}
\bigskip

$^a$ from de Vaucouleurs et al. 1991. PA is the position angle of the 
optical major axis; $B_{\rm T}^0$ is the total B magnitude, corrected for 
galactic and internal extinction; $R_{\rm e}$ is the 
effective radius, and $b/a$ is the minor to major axes ratio.

$^b$ distance from Fabbiano et al. (1992), who adopt a Hubble constant of 50 
km s$^{-1}$ Mpc$^{-1}$.

$^c$ total B-band luminosity \lb, derived using the indicated distance and 
$B_{\rm T}^0$. 

$^d$ central stellar velocity dispersion from McElroy (1995). 

$^e$ Galactic neutral hydrogen column density from Stark \etal (1992).

\end{table*}

\section { General characteristics  of NGC~3923}

The main characteristics of NGC~3923 are summarized in Table 1.
This is the dominant galaxy of a small group of five objects (LGG255, Garcia 
1993); the other members are one elliptical (NGC~3904), one lenticular and two 
spirals. The velocity dispersion of the group is 184 km s$^{-1}$ (Huchra \&
Geller 1982). No extended X-ray emission has been detected other than that
associated with the galaxy, from a possible intragroup medium (Forman et al. 1985).
NGC~3923 is one of the best known examples of an elliptical
galaxy with shells (e.g., Carter et al. 1998). In NGC~3923 the shells are expected to be the result 
of two possible processes: a low-velocity merger with a less massive galaxy 
which is eventually disrupted, on a long time scale of several Gyrs (Quinn 
1984);  or a weak interaction, which took place a couple of Gyrs ago, with a 
passing galaxy of much smaller mass (Thomson 1991).
NGC~3923 shows no rotation on its major axis (e.g., Pellegrini et al. 1997), and small minor axis rotation (Carter et al. 1998),
which is consistent with either the presence of a kinematically  decoupled
core, established during a merger, or triaxiality (Franx et al. 1989). 
The modest content of cold gas and dust in NGC~3923 
is consistent with a merger with a dwarf spheroidal; this would also explain
the absence of radio activity, or of enhanced star formation (Forbes 1991). In case, any 
enhanced activity is expected to have occurred on a shorter time-scale than
that of the long-lived shells.
In all other respects except the presence of shells, NGC~3923 appears to be
an ordinary elliptical galaxy, whose projected image on the sky is quite 
flat.

\section {X-ray Data Analysis}
NGC~3923 was observed by two Narrow Field Instruments on
board the {\it BeppoSAX} satellite (Boella et al. 1997a):
the Low Energy Concentrator Spectrometer
(LECS) and the Medium Energy Concentrator Spectrometer (MECS).
The journal of the observation is given in Table 2.
LECS and MECS are made of grazing incidence telescopes with 
position sensitive gas scintillation proportional counters in their focal
planes. The MECS, which consists of three equal units, has a field of
view of 56 arcmin diameter, works in the range 1.5--10 keV, with an
energy resolution of $\sim 8$\%,
and moderate spatial resolution of $\sim 0.7$ arcmin FWHM at 6 keV
(Boella et al. 1997b). The total effective area of the MECS 
is comparable to that of the 2 GIS units on board {\it ASCA}.
The LECS is sensitive also at softer energies (over 0.5--10 keV),
has a field of view of 37 arcmin diameter, an energy resolution 
a factor of $\sim 2.4$ better than that of the {\it ROSAT} PSPC, but 
an effective area much lower (between a factor of 6 and 2 lower,
going from 0.3 to 1.5 keV; Parmar et al. 1997).

The cleaned and linearized data have been retrieved from the {\it BeppoSAX} 
Science Data Center archive, and later reduced and analysed using the standard 
software (XSELECT v1.3, FTOOLS v4.0, IRAF-PROS v2.5, and XSPEC v10.0). For 
the MECS, the event file made by merging the data
of the 3 MECS units, properly equalized, has been used.

\begin{figure*}
\vspace{-23cm}
\vspace{5cm}\psfig{figure=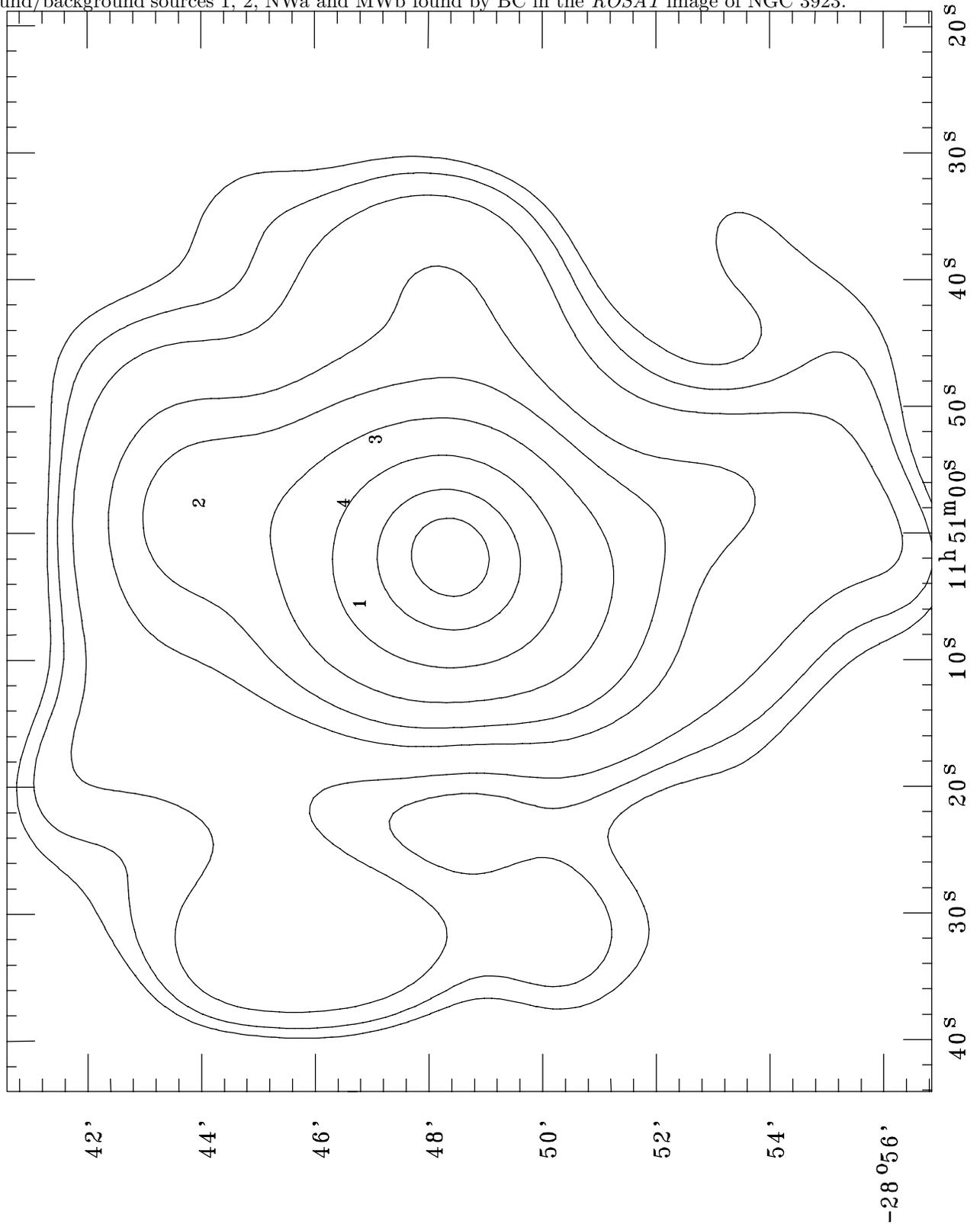,bbllx=80pt,bblly=150pt,bburx=480pt,bbury=700pt}  
\vspace{0cm}
\caption[]{The MECS image of NGC~3923, smoothed with a gaussian of $\sigma
=1^{\prime}$. Contours plotted correspond 27,
 30, 33, 39, 45, 55, 70, 85, 95 \% of the peak intensity. 
The numbers 1, 2, 3, 4 mark respectively the positions of the
foreground/background sources 1, 2, NWa and MWb found by BC in the {\it ROSAT}
image of NGC~3923.}
\vskip 23cm
\end{figure*}

\begin{table*}
\caption[]{ Observation Log}
\begin{flushleft}
\begin{tabular}{llllllllll}
\noalign{\smallskip}
\hline
\noalign{\smallskip}
 Date        & \multicolumn{2}{l}{Exposure$^a$ (ks)}  &\  & 
\multicolumn{2}{l}{Count Rate$^b$ (ct/s)}             &\  &
\multicolumn{2}{l}{Region$^c$} \\
\cline{2-3}\cline{5-6}\cline{8-9}
             & LECS & MECS  & & LECS & MECS &  & LECS & MECS  \\
\noalign{\smallskip}
\hline
\noalign{\smallskip}
1997 Jan 29  & 16.08& 43.51 & & 0.01$\pm 0.0015$ & 0.01$\pm 0.0010$ &  & $R=8^{\prime}$ & $a=8^{\prime},
\,\,b=6^{\prime}$ \\ 
\noalign{\smallskip}
\hline
\end{tabular} 
\end{flushleft}
\bigskip

$^a$ On-source net exposure time. The LECS exposure time is considerably reduced
with respect to the MECS one, because the LECS can operate only when the
spacecraft is not illuminated by the Sun.

$^b$ Background subtracted source count rates, with photon counting statistics 
errors, within the regions specified in the last column.

$^c$ The extraction region is a circle for the LECS data and an ellipse 
of semi-axes $a$ and $b$ for the MECS data.

\end{table*}

\subsection{Spatial Analysis}

In Fig. 1 a contour plot of the MECS image is shown. The counts in this
image belong to pulse invariant gain-corrected
 spectral channels 29--218 ($\sim$1.7--10 keV).
The data have been smoothed with
a gaussian of $\sigma =1^{\prime}$. An extended source is
clearly visible, of extension larger than the PSF, and covering at least
8 optical effective radii. The X-ray
emission shows roughly the same position angle of the optical emission.
The background subtracted radial profile obtained from the MECS data
over 1.7--10 keV is plotted in Fig. 2. 

The X-ray source extent needs to be estimated precisely, in order 
to determine the size of the region from which to extract the counts for the
spectral analysis, and the estimate of the flux.
We consider the source boundary to be the point at which the total
(i.e., source$+$background) X-ray surface brightness 
 flattens onto the background level. 
The background is estimated from blank fields event files\footnote{
This choice was motivated by the fact that at radii $>10^{\prime}$ the energy dependent
vignetting of the mirrors is not negligible.
In the position of the galaxy instead, i.e., on-axis within radii of $10^{\prime}$, 
the vignetting effect is negligible, well within the uncertainties in the 
profile.
}, accumulated
on five different pointings of empty fields, and using
extraction regions corresponding in size and position to those of the source.
In addition to the inspection of various surface brightness profiles (e.g.,
the azimuthally averaged one, and also those along the major and minor axes),
the precise source extent was finally established also through 
the requirement that within it the S/N of the background-subtracted counts
keeps at a high value, not much lower than that of the central
regions of the galaxy.
This produced an extraction region of an ellipse for the MECS data, of
semimajor axis of $8^{\prime}$ and semiminor axis of $6^{\prime}$ (Table 2).
The PSF of the LECS is a strong function of energy, and it is 
broader than that of the MECS below 1 keV, while it is similar to that
of the MECS above 2 keV.
A radius of $8^{\prime}$ is suggested to encircle all the
photons of soft sources (http://www.sdc.asi.it/software/cookbook).
This radius turned out to be optimal in our case, satisfying both the
requirement of being the radius at which the brightness profile flattens onto the
background level, and that of a high S/N.
These different extraction regions for the LECS and the MECS images encircle
 the same fraction of the source photons, in a given band common to the 
two instruments, as verified later during the spectral analysis.

Analyzing the {\it ROSAT} PSPC image of NGC~3923, BC detected 
a few foreground/background sources; those falling within the extraction
region are displayed in Fig. 1. The nature of these sources is unknown. 
Sources called "1", "NWa", and "NWb" by BC cannot be 
distinguished in the MECS image, while this image looks quite aligned to the 
North-South due to source "2". By deriving the net counts in each one of four quadrants,
obtained by dividing the extraction region ellipse with its major and minor 
axes,
it turns out that the quadrant comprising sources "1","2", 
and "NWb" has roughly 60$\pm$20 net counts more than the average
of the others. This is a contribution of $\sim 14$\% to the MECS net counts. 
The quadrant with source "NWa" shows no significant net counts enhancement.
By inspecting the surface brightness profile in a stripe across source "2", 
a contribution of $\sim 16$ net counts is estimated from this source; this is
likely to be responsible for the jump at a radius of $260^{\prime\prime}$
in the azimuthally averaged surface brightness profile of Fig. 2. 
Further implications coming from the presence of these sources 
for the spectral analysis are discussed in Sect. 4.

\begin{figure}
\vspace{0cm}
\hspace{0cm}\psfig{figure=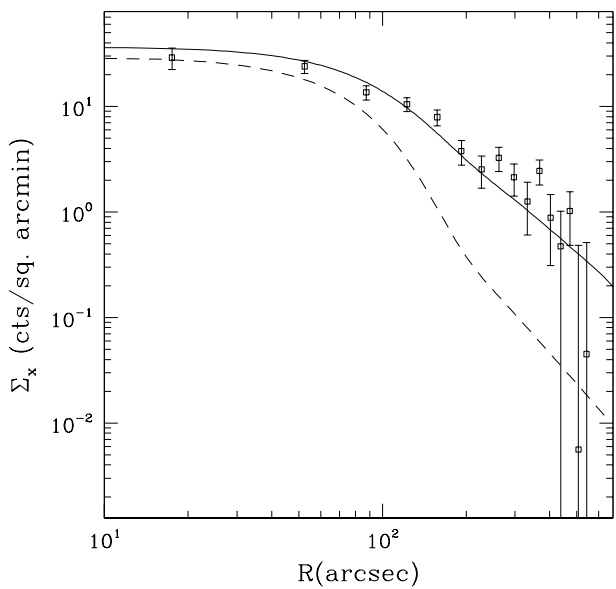,width=8.8cm,bbllx=80pt,bblly=50pt,%
                   bburx=480pt,bbury=600pt}
\vspace{0cm}
\vskip -4.truecm
\caption[]{ Background subtracted azimuthally averaged surface brightness profile obtained
from the MECS data over $1.7-10$ keV).
Also plotted is the PSF of the MECS for $E=5 $ keV (dashed line), and the convolution
of the optical profile with the MECS PSFs calculated at the source photon energies 
(solid line; see Sect. 4.1). One pixel size is $8^{\prime\prime}$.
}
\end{figure}

\subsection {Spectral Analysis}

Within the source regions determined as described above, we obtained the
net count rates given in Table 2, and finally extracted the LECS and MECS 
spectra. 
Spectral channels 
covering the energy ranges 0.3--10 keV, and 1.7--10 keV respectively 
for the LECS and MECS have been used.
The original channels have been grouped into larger bins, adequately filled for 
applicability of the $\chi^2$ statistic to assess the goodness of 
a fit (Cash 1979). The data have been compared to models, convolved
 with the instrumental and mirror responses, using the $\chi ^2$ minimization 
method. The spectral response matrices released in September 1997
have been used in the fitting process.
We fitted the models simultaneously to the LECS and MECS spectral data.

Given the discussion of Sect. 1, the basic models used for the fits 
are the bremsstrahlung model, and two models describing the thermal
emission of an optically thin hot plasma, both from continuum and lines:
the standard Raymond-Smith model (hereafter R-S, Raymond \& Smith 1977), and the  MEKAL model.
The latter is a modification of the original MEKA model (Kaastra \&
Mewe 1993) where the Fe-L shell transitions have been recalculated recently
according to the prescriptions of Liedahl et al. (1995).
The abundance ratios of the heavy elements cannot be constrained by the 
modeling, because of the poor statistics. So, we assume that the
abundance ratios are solar, as accurately determined by Matsushita et al. (1998) for
 the ISM of the X-ray bright galaxy NGC 4636, using a 
very long {\it ASCA} exposure.
The solar abundances are those given by Feldman (1992; e.g., the abundance of
iron relative to hydrogen is 3.24 $10^{-5}$ by number, also known as meteoritic
abundance).
The models are modified by photoelectric absorption of the X-ray photons  
due to intervening cold gas along the line of sight, of column density $N_H$;
we take absorption cross sections according to Baluci\'nska-Church \&
McCammon (1992).
The results of the spectral analysis are presented in Tables 3 and 4.

\begin{table*}
\caption[]{ One-component spectral fits}
\begin{flushleft}
\begin{tabular}{ lllllll rl }
\noalign{\smallskip}
\hline
\noalign{\smallskip}
 Model     & $10^{-20}N_{\rm H}\, ^a$ & $kT$ & Z             & $\chi^2$ & 
$\nu \, ^b$ & $\chi^2/\nu$ \\
           & (cm$^{-2}$)         & (keV)& (Z$_{\odot}$) &      &       & \\
\noalign{\smallskip}
\hline
\noalign{\smallskip}
\tbsp
Bremsstrahlung & 1.  & 2.9 (2.0--5.4)    &                      &  29.1   &  
 24      &   1.2 \\
               & 6.4       & 3.1 (2.0--5.3)    &                &  33.1   &
 25      &   1.3 \\
\hline
\tbsp
R-S            & 1.  & 3.2 (2.1--5.4)    & 0. (0.--1.8)         &  28.9   &
 23      &   1.3 \\
               & 1.  & 3.0 (2.2--4.4)    & 1.0 (fixed)          &  30.2   &
 24      &   1.3 \\
               & 6.4       & 3.4 (2.0--5.4)& 0.02 (0.--2.2)     &  33.0   &
 24      &   1.4 \\
               & 6.4       & 3.0 (2.1--4.3)& 1.0 (fixed)        &  34.0   &
 25      &   1.4 \\
               \hline
\tbsp
MEKAL          & 1.  & 3.2 (2.1--5.3)    & 0.07 (0.--1.8)       &  29.0   &  
 23      &   1.3 \\
               & 1.  & 3.1 (2.2--4.0)    & 1.0 (fixed)        &  30.1   &  
 24      &   1.3 \\
               & 6.4       & 3.4 (2.0--5.4)    & 0.05 (0--2.6)  &  33.2   &
 24      &   1.4 \\
               & 6.4       & 3.0 (2.1--4.3)    & 1.0 (fixed)        &  33.9   &
 25      &   1.4 \\
\hline
\tbsp
Cooling        & 1.        & 5.7 (3.4--9.6)    & 0.45 ($>0.02$)&  21.6   &
 23      &   0.9 \\
flow$^c$       & 1.        & 4.7 (2.9--7.5)    & 1.0 (fixed)        &  22.3   & 
 24      &   0.9 \\
               & 6.4       & 4.6 (2.9--7.5)    & 0.60 ($>0.05$) &  24.8   &
 24      &   1.0 \\
               & 6.4       & 4.6 (2.9--7.5)& 1.0 (fixed)        &  25.0   &
 25      &   1.0 \\
\noalign{\smallskip}
\hline 
\end{tabular} 
\end{flushleft}

$^a$ When in this column $N_{\rm H}=6.4\times 10^{20}$ cm$^{-2}$, it has been kept at this 
value (the Galactic one) during the fit. When $N_{\rm H}$ is a free parameter 
of the fit, it is
constrained to lie in the interval $10^{20}-10^{21}$ cm$^{-2}$, and its
"best fit" value always turns out to be the lower boundary of the interval.

$^b$ Number of degrees of freedom in the fit.

$^c$ In this case $kT $ in the third column gives the maximum temperature 
from which the gas is cooling.

\smallskip
The ranges of values between parentheses indicate the 90\% confidence
range of variation for one interesting parameter.

\end{table*}

\bigskip

\begin{table*}
\caption[]{Two-component spectral fits}
\end{table*}

\subsubsection{One-component models}

We first tried to fit one-component models to the data (Table 3). The probabilities of
exceeding the reduced $\chi^2_{\rm min}$ are quite low (they range
from 0.1 to 0.2) but the fits are
formally acceptable. At the best fit $N_H$ tends to be lower than the Galactic value;
 the fits are not worsened significantly  when keeping $N_H$ fixed at the
Galactic value. All the one-component models give fits of comparable
quality, and a temperature around 3 keV.
 The abundance at the best fit is extremely low, both
for the MEKAL and the R-S models, but it is just constrained to be
$<(2-3)\, Z_{\odot}$ by the data. So, we have also performed some fits with
the abundance fixed at the solar value. 
These findings are in partial agreement with those obtained from {\it ASCA}-SIS
 data over (0.5--5) keV by BF:
when keeping $N_H$  fixed at the Galactic value, the single-component MEKAL
fit gives them an abundance $Z=0.07\,Z_{\odot}$ close to that found by us,
but a temperature of only $kT=0.64$ keV. This could reflect the different 
extraction regions used for the spectrum, if the emission is much softer at 
the center; or could be the result of different spectral sensitivities
between the {\it BeppoSAX}-MECS and the {\it ASCA}-SIS, from which the
best data come in the cases of the two satellites.
The spectral models derived from {\it ASCA} data are mostly
constrained by the SIS data near 1 keV (BF), while those  derived from
{\it BeppoSAX} are mostly constrained by the MECS data
at energies above that region.  

\subsubsection{Multi-component models}

We then tried to fit multi-temperature models, to check whether the quality
of the fits can be improved. Since the R-S model gives results
substantially equal to those given by the MEKAL model, for these data,
we give in the
following only the results obtained by using the more updated MEKAL model.

First we tried the fit with a cooling flow model, in which the cooling is
isobaric, and the emissivity is as described in Johnstone et al. (1992); this model is made 
by the superposition of many thermal components, each described 
by a MEKAL model at a certain temperature. The results are given in Table 3; the fits are improved
(the associated probability is 0.4--0.5).
The upper temperature from which the gas cools 
is very high, $\sim 4.7$ keV; this is also the temperature at which the emission
measure peaks (the emission-weighted temperature is somewhat lower).
This modeling reveals that hard emission is present at a significant
level, so that the distribution of the temperatures is forced to extend to
high values (``high'' for the temperatures expected in a galaxy cooling flow).
The upper temperature values found here are higher
than that given by BF ($kT_{max}=1.13$ keV)
for a fit with the same model, 
with $N_H$  fixed at the Galactic value, and $Z=0.11\,Z_{\odot}$ resulting
from the fit. The reasons for this discrepancy are likely the same given
above for the one-component case.
In agreement with BF we
find that the abundance at the best fit (0.6 $Z_{\odot}$) is raised with respect to that 
obtained from the one component MEKAL fit.

Then we tried the  coupling of two thermal components (MEKAL+bremsstrahlung,
and MEKAL+MEKAL) with the same $N_H$ (Table 4).  The quality of the fits
is much improved now, and the probability of exceeding $\chi^2_{\rm min}$ is 
very large ($>0.9$). 
As before, allowing $N_H$ to be free does not 
significantly improve the fits; when $Z$ is fixed at 1 $Z_{\odot}$, $N_H$ at
the best fit is reasonably close to the Galactic value. 
In Fig. 3 we show the LECS and MECS data 
together with the two-component model which gives the best fit.
More in detail, the results obtained by fitting with two thermal components
are as follows: 

a) MEKAL + bremsstrahlung models: the first
turns out to describe 
a soft component, of temperature \ts$\approx$0.4 keV, and the second
describes a hard component of temperature \th\sp between 6.2 and 7.6 keV.
\th \sp is higher for a lower
metallicity of the soft component. The abundance of the soft
component is again very subsolar at the best fit (0.16 $Z_{\odot}$), 
and not constrained by the data.

b) Two MEKAL components:  the results are similar.
The abundance of the hard component turns out to be practically zero
at the best fit; that of the soft component is also very low (0.1 $Z_{\odot}$),
but again $Z$ is just constrained to be $>0$.
At the best fit the temperatures are \ts\sp around 0.4 keV and \th\sp around 5 or 8 keV,
 depending
on the abundance, fixed at the solar value or allowed to vary freely
 respectively. As before
the temperature of the hard component is higher for a lower
metallicity, but the effect is more pronounced now.
In this kind of fit
 BF find temperatures close to ours (\ts=0.55 keV, and \th=4.2 keV), but
$Z=2.0\,Z_{\odot}$ ($>0.1\,Z_{\odot}$ at 90\% confidence), with $N_H$ kept at 
the Galactic value\footnote{As for the previous two-component fits,
the first best fit value of the abundance given by XSPEC is around solar, but
an accurate exploration of the parameter space finds the true minimum $\chi^2$
at a much lower abundance value. This behavior is somewhat opposite to that
found by BF when fitting the {\it ASCA}-SIS data, which could be explained 
by the lower spectral sensitivity around 1 keV of the {\it BeppoSAX}-LECS.}. 
 
\begin{figure}
\vspace{0cm}
\hspace{0cm}\psfig{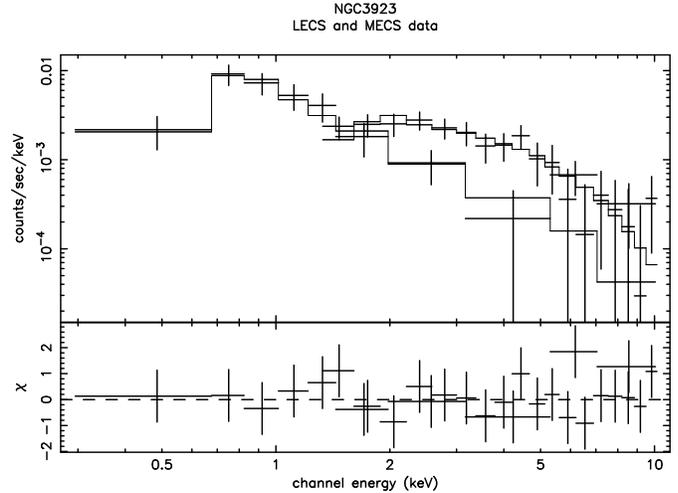}
\vspace{0cm}

\caption[]{The LECS and MECS spectral data with the best fit 
MEKAL + bremsstrahlung model with solar abundance and $N_H$ fixed at the 
Galactic value, shown with thinner lines, and the residuals from the fit.
}
\end{figure}

\smallskip
\par\noindent
Summarizing, all the two-component models give \ts\sp around 0.4 keV,
and \th\sp around 6--8 keV, except when the hard component
is described by a MEKAL model with solar abundance (then \th$\approx 5$ keV).
The latter case is also the one giving the worst fit among all the 
two-component models ($P=0.85$). Heavy element abundances 
at the best fit tend to be very subsolar, except for the cooling flow model.
The effect of imposing an abundance higher than that at the
best fit is to produce a lower temperature of the hard component.
This effect is particularly pronounced
when fitting with two components (respectively \th\sp decreases of $\sim 1.5$ 
to $\sim 3$ keV, using the bremsstrahlung and the MEKAL model for the hard component).

\subsection {X-ray Fluxes and Luminosities}

In Table 4 the absorbed fluxes in the (0.5--4.5) and (0.5--10) keV bands
are  given; unabsorbed fluxes are about 30\% higher for the soft component,
and 5 to 9\% higher for the hard one, depending on the band.
An average best fit value of the total absorbed flux $F(0.5-4.5$ keV) is $\sim 8\times 10^{-13}$ erg cm$^{-2}$ s$^{
-1}$, and $F(0.5-10$ keV) $\sim 10^{-12}$ erg cm$^{-2}$ s$^{-1}$. 
The ratio between the absorbed fluxes in the soft and in the hard components
in the (0.5--4.5) keV band is $\sim 0.93$, close to the
value of 0.99 found by BF in the (0.5--5) keV band. This ratio actually is 
$\sim 0.8$ when $Z$ is 
kept at $1\,Z_{\odot}$ (and \th$\sim 6$ keV), and $\sim 1.1$ when $Z\sim 0$ 
(\th$\sim 8$ keV).
An average value for the ratio between the absorbed fluxes in the soft and in the hard 
component in the (0.5--10) keV band is $\sim 0.61
$. Again it is lower (i.e., $\sim 0.5$) when $Z$ is kept at $1\,Z_{\odot}$,
and higher ($\sim 0.7$) when $Z$ is very subsolar. 
In the (0.5--4.5) keV band, the soft component unabsorbed flux is 52\% of the
total, while the hard component amounts to $\sim 57$\% of the 
total (0.5--10) keV unabsorbed flux.

Adopting the distance in Table 1, the fluxes can be converted into the
following values of the luminosities in the (0.5--4.5) keV band: an average
value of the soft component absorbed luminosity is $L_X=7.8\times10^{40}$ erg 
s$^{-1}$ (unabsorbed $L_X=1.0\times 10^{41}$ erg s$^{-1}$),
and of the hard component is $L_X=8.2\times 10^{40}$ erg s$^{-1}$
(unabsorbed $L_X=8.9\times 10^{40}$ erg s$^{-1}$).

\section { Discussion}

\subsection {Origin of the X-ray emission in NGC~3923}

The spectral analysis over the (0.5--10) keV band showed that
the superposition of two thermal 
components of \ts$\approx 0.4$ keV, and \th$\approx 6-8$ keV, is the most 
reasonable representation of the spectral data. Whether the softer component
is actually a multi-temperature component, such as a cooling flow, cannot be
investigated with these data. Another caveat concerns the hard component, 
because, as seen in Sect. 3.1,  BC detected a few point sources over the 
{\it ROSAT} PSPC image of NGC~3923. 
Were all these sources background AGNs, they could 
 harden the spectrum of NGC~3923, and alter the estimate of 
the hard flux from NGC~3923. Given the number counts involved (see Sect. 3.1),
 the effects on the derived spectral parameters
are expected to be within their uncertainties; in particular,
there could be a spuriuos increase in the hard flux of $\lsim 14$\%. 

The value of \ts\sp suggests as origin of the
soft emission a hot gas, that  comes from the accumulation of the stellar mass
loss during the galaxy lifetime. The value of \th\sp is close to that found
for the hard emission of the low \lx/\lb\sp galaxy NGC~4382,
based on {\it ASCA} data (i.e., $\sim 6$ keV; Kim et al. 1996). Following
 Fabbiano \etal (1992, 1994), these authors suggest for the origin
of the hard component the integrated emission of LMXBs, an interpretation similar
 to that given for the X-ray emission of the bulges of early-type
spirals (see Sect. 1).  In fact the spectra of LMXBs
can be described by a thermal bremsstrahlung model with $kT\sim 5$ keV
(van Paradijs 1998), the spectrum of M31 can be fitted by
bremsstrahlung emission at a temperature $kT>3$ keV (Fabbiano et al. 1987), and those of bulge-dominated spirals can be fitted with 
bremsstrahlung at $kT\gsim 5$ keV (Makishima et al. 1989).
We now examine more in detail the suggested origins for the
two components of the X-ray emission of NGC~3923.

\subsubsection{Origin of the soft component}

How does \ts\sp compare with the possible average temperature of a gas flow in NGC~3923?
A simple estimate of the kinetic temperature of the stars in NGC~3923
gives $kT=\mu m_p\sigma^2=0.36$ keV, when using the central stellar velocity
dispersion in Table 1, and $\mu =0.6$. This is close to \ts\sp found by the 
modeling, i.e., \ts=0.39--0.45 keV, with
 some room left for other heating mechanisms in addition to the thermalization
of the stellar random motions. These are especially needed
when considering that 0.36 keV is the {\it central}
temperature of the stars, and that this decreases outward, while \ts\sp would be
the emission-weighted temperature of the hot gas. Possible heating mechanisms 
are supernova heating (from type Ia supernovae, hereafter SNIa) and/or compressional 
heating operated by the gravitational field\footnote{Compression can be
operated also by an external medium, but this has not been observed around 
NGC~3923. Another more exotic possibility is Compton heating produced by 
accretion onto a supermassive black hole at the galaxy center
(Ciotti \& Ostriker 1997).}. 
Compressional heating is produced when the hot ISM is in a global inflow.
The amount of the emission in the soft component ($\sim 10^{41}$ erg s$^{-1}$)
is actually  quite lower than that typical of
a global inflow, in a galaxy of an optical luminosity as high as that of 
NGC~3923.
For example, when $L_B\approx 2\times 10^{11}$\lsun, \lx\sp of many times 
$10^{41}$ erg s$^{-1}$ in the (0.5--4.0) keV band, and hot gas masses\footnote{
The precise amount of hot gas can be calculated when the emissivity profile
is known (i.e, in the hypothesis of spherical symmetry, $\epsilon (r)=
\Lambda(T,Z)\;n(r)^2$, where $r$ is the spatial radius, $\Lambda$ is the 
cooling function and $n$ the gas density), and this requires an accurate
surface brightness profile, temperature profile, and abundance estimate,
 for the hot gas. Under the hypotheses of solar
abundance, isothermality, and a standard $\beta$-model for the hot gas 
distribution (CFT), one has $M_{gas}=5.0\times 10^9\sqrt{L_X/10^{41}}=5.0\times
 10^9 $\msun\sp for NGC~3923.
This is estimated to be accurate to factors of 2--4.} of a few
times $10^{10}$\msun, are
 predicted by the steady state cooling flow models, for various combinations
 of dark matter and SNIa rates (e.g., Sarazin \& White 1987, Bertin \& Toniazzo 1995). 
Besides eliminating SNIa's, various kinds of reductions of the high \lx
\sp values in the framework of the cooling flow scenario have been suggested: 
by assuming, at fixed \lb, reductions in the 
stellar mass loss rate, or in the efficiency of its thermalization, or a higher
efficiency of thermal instabilities in the hot gas (Sarazin \& Ashe 1989,
Bertin \& Toniazzo 1995), or by including the effect of the rotation of
the galaxy (Brighenti \& Mathews 1996), or of a lower abundance (Irwin \&
Sarazin 1998). None of these effects has proved to be able to reduce \lx\sp by
a large factor.
\lx\sp up to $10^{41}$ erg s$^{-1}$, and gas temperatures higher than the
stellar kinetic temperatures, are instead predicted for a galaxy like NGC~3923
from a gas  that is experiencing an outflow or a partial wind; these 
can be driven by SNIa explosions at a rate comparable to the most recent optical estimates of
Cappellaro et al. (1997) (Pellegrini \& Fabbiano 1994, Pellegrini \& Ciotti 
1998). In addition, within
this framework, global energy considerations and 
two-dimensional simulations showed that in general the flattening of the galaxy favors the loss of
gas, while rotation has a minor role (Ciotti \& Pellegrini 1996; D'Ercole \&
Ciotti 1998). This could be an explanation for the
absence of a global inflow in NGC~3923, which is a considerably flat galaxy
with no rotation (Sect. 2).
A detailed modeling of the structure of NGC~3923, and hydrodynamical
simulations of the hot gas behavior specific for this galaxy,
are required for a definite answer about the gas flow state.
There is a potential problem with a scenario that involves a substantial
heating from SNIa's.
The abundance of the soft component is not constrained by the {\it BeppoSAX}
data for NGC 3923, but is extremely low at the best fit. 
Low abundances for the hot gas are
not expected in presence of SNIa explosions, yet they have almost always 
resulted also from the analysis
of {\it ASCA} data; there seems 
even to be a trend of decreasing abundances with decreasing
\lx/\lb\sp (Matsushita 1998). 
An unsolved puzzle is represented at present by the discrepancy between
the low hot gas abundances and the abundances in the stellar 
mass loss which feeds the gas (but this discrepancy is narrowing for an increasing number of galaxies,
after accurate re-analyses; Matsushita 1998, BF, Buote 1998), eventually further increased
by the metals in the ejecta of SNIa's, which are seen to explode in E/S0s. Various solutions
have been suggested (Arimoto et al. 1997, Fujita et al. 1996), but none has
been recognized as the final one yet.
We note that in NGC~3923 the stellar central
iron abundance is [Fe/H]=0.2, or 1.6 solar, and that the mean abundance is 
likely a factor of 2 lower (the central Mg$_2$ value, from Faber et al. 
1989, has been converted into central and average stellar iron
abundance following the detailed prescriptions of Arimoto et al. 1997).
If we adopt the abundance value of $Z=2.0\,Z_{\odot}$ for the hot gas
found by BF (Section 3.2.2), in this galaxy there is room for enrichment
by SNIa's exploding at the rate of Cappellaro et al. (1997). 

Another possibility to explain the gas mass content of
NGC~3923 is that a substantial amount of hot ISM was lost
as a consequence of the episod of interaction or merger which
is at the origin of the system of shells shown in the optical. 
Actually, the very detailed modeling of the shell 
formation that has been made for this galaxy, plus observations
of the galaxy colors and ISM content at other wavelengths (see Sect. 2),
established that the interaction or merger involved a small galaxy, devoid of 
gas, and that significative star formation (in the form of a starburst with 
supernova explosions that could have heated the gas) did not take place.
It could be, though, that the gas flow is very sensitive to 
perturbations in the potential, and that even small perturbations can
help a significant portion of the hot ISM to escape the galaxy.
Numerical simulations are needed to test whether this was the case for 
NGC~3923.

\subsubsection{Origin of the hard component}

\th\sp is in good agreement with that of bulge-dominated spirals; what about the
amount of the hard emission in NGC 3923?
CFT had estimated the luminosity  of the integrated contribution
of LMXBs in the (0.5--4.5) keV band ($L_{dscr}$)
by scaling it from the emission of the bulge of M31. By assuming a linear relation with \lb,
they had obtained log $L_{dscr}$ = 29.6 + log \lb, where \lb\sp is in \lsun, 
and had estimated $L_{dscr}$ for any given galaxy to scatter by about a factor
 of 3 about this relation, since this is the observed scatter in the X-ray to optical
luminosity ratio for subclasses of spiral galaxies.
This relation has been recently confirmed (both in shape and normalization)
using {\it ASCA} data by Matsumoto et al. (1997). 
The present analysis gives, in the (0.5--4.5) keV band, log $(L_{X,hard}/L_B)
\approx 29.65$, i.e., $L_{X,hard}$ is close to the value predicted
 for $L_{dscr}$ by CFT (it is just 12\% higher, so well within the quoted 
uncertainties).
An interpretation in terms of stellar sources of the hard emission can be 
judged  also by inspecting Fig.~2, where the MECS surface brightness profile is
compared with the distribution of optical light. The V-band profile of
 NGC~3923 has been derived from Kodaira et al. (1990); since this extends out to a 
radius of $262^{\prime\prime}$, it has been extrapolated out to the radius
of the X-ray emission with a fit.
The optical profile has then been convolved with the MECS response, appropriate
 for the distribution of the counts in the different energy channels.
A good agreement between the  profile  over (1.7--10) keV and the convolved optical profile
is found, which would support the hypothesis of the origin of the hard emission
in stellar sources. We note here that, consistently with the finding of an 
amount of hard emission 12\% larger than the predicted $L_{dscr}$, and
with the estimate of a possible spurious contribution up to 14\% of the MECS 
counts from hard foreground/background sources, some 
excess of hard emission is also shown by the X-ray
profile, with respect to the convolved stellar profile, at 
radii $\gsim 260^{\prime\prime}$, i.e., for $R>5 R_e$ (the excess at
$260^{\prime\prime}$ is likely due to a foreground/background source,
see Sect. 3.1).
We cannot give a great significance to the detailed shape of the X-ray
profile, because of the MECS moderate spatial resolution (Sect. 1); we just note 
that this hard excess cannot be produced by hot gas, 
neither belonging to the galaxy (at \ts=0.4 keV),
nor to a possible intragroup medium, because the low value of the velocity dispersion 
of the group (Sect. 2) corresponds to a kinetic temperature that is even lower 
than \ts.
An X-ray profile of NGC~3923 with a superior spatial resolution has been derived in the 
soft {\it ROSAT} band (0.4--2) keV by BC. For
radii $>10^{\prime\prime}$ this is in good agreement with 
the de Vaucouleurs  extrapolation 
of the R-band profile within $103^{\prime\prime}$, while it is flatter than the
optical one for radii $< 10^{\prime\prime}$. BC conclude that not all the hard emission is to
be attributed to stellar sources, while some fraction of it could come from
another phase of the hot gas. In line with the modeling done to interpret the 
{\it ROSAT} data of another galaxy which showed an X-ray profile centrally flatter than 
the optical one (NGC~4365; Pellegrini \& Fabbiano 1994), we suggest the 
possibility that the hot gas has quite an extended 
distribution in the central regions, i.e.,  flatter than the optical one 
within 10 arcsec (consistently with what can be derived also by the {\it BeppoSAX}
data, BC estimate that just 35\% of the total 0.4--2 keV emission is due to a
hard component; so this cannot fully determine the total shape).
The hypothesis of some peculiarities in the hot gas distribution can be 
supported also by the consideration of the past galaxy history, where a merging
occurred.

\subsection {The nature of the X-ray emission in medium and low \lx/\lb\sp galaxies}

A fundamental diagnostic of the X-ray emission from early-type galaxies is the
\lx--\lb\sp plane. The large scatter in \lx\sp of more than two orders of
magnitude at fixed \lb\sp shown by this plane is
 not an artifact of distance errors
[see Pellegrini \& Ciotti (1998) for a more detailed discussion].
The explanation of this scatter is a largely varying
quantity of hot gas within the galaxies (e.g., Matsumoto et al. 1997), but it is still 
a controversial issue how these variations are established.  
Are they fundamentally a consequence of environmental differences, or of 
different dynamical phases for the hot gas flows (provided that it was not possible 
to reproduce the observed scatter with various adjustments to the cooling flow scenario,
see Sect. 4.1.1)?
The first hypothesis can affect only galaxies in clusters or groups; actually this is the case 
for the majority of E/S0s.
Then accretion of external gas can explain the extremely
X-ray bright objects (Renzini et al. 1993, Mathews \& Brighenti 1998), while
in the X-ray faint ones the hot gaseous halos should have been stripped 
by ambient gas, if it is sufficiently dense, or in encounters with
other galaxies (White \& Sarazin 1991). It is not clear yet 
whether the primary stripping agents would be other galaxies or the ambient gas.
The effectiveness of the stripping by an ambient gas has been
explored theoretically, and  it turned out to depend largely on various 
factors (shape of the orbit, velocity and internal dynamics of
the galaxy, density of the environment),
for which the observed range is wide (e.g., Portnoy et al. 1993).
Observationally, evidence of stripping by the intracluster medium is 
the famous plume shown by the hot halo of the Virgo elliptical M86.
{\it ROSAT} though showed that the sample of early-type galaxies of the Coma cluster,
that is richer than Virgo, has the same average \lx/\lb\sp as that of Virgo
(Dow \& White 1995); but that  instead the X-ray luminosities are on average
lower in the rich cluster A2634 (Sakelliou \& Merrifield 1998).
In the Pegasus I group and in the poor cluster
Cancer A, where a medium has been detected, the X-ray image shows also many 
clumps that could be the X-ray halos from individual galaxies 
 with a `normal' \lx/\lb\sp (Trinchieri et al. 1997). 
For what is concerning the interactions
among galaxies, observationally there is an indication that lower \lx/\lb\sp 
galaxies occur across the whole range of galaxy densities, while the higher 
\lx/\lb\sp ones are mostly confined at low densities (Mackie \& Fabbiano 1997). 
Theoretically, galaxy interactions could  produce some scatter in 
\lx/\lb\sp as follows: group-dominant
ellipticals may acquire dark matter and hot gas by mergers or tidal
interactions early in their evolution, and then become very X-ray bright; the
other E/S0s, in which the gas is in a global inflow, may be tidally truncated in their 
dark matter and hot gas halo,
at different radii, and so end up with different sizes and different \lx/\lb\sp
(Mathews \& Brighenti 1998).
The problems with explaining the \lx/\lb\sp plane only with environmental 
factors are that: 1) low or medium \lx/\lb\sp values are also shown by 
galaxies that do not reside in a high density medium [e.g., NGC 5866
(Pellegrini 1994), NGC 3923], and by galaxies that reside in a region where
the galaxy density is not particularly high, and where also galaxies of
high \lx/\lb\sp are found (Mackie \& Fabbiano 1997);
2) the effect of merging and tidal interaction on the
hot gas flow (in various dynamical states) is still quite conjectural, as
is the evolution of the hot gas in galaxies that undergo 
these phenomena. The only models available so far,
those of Mathews \& Brighenti (1998), are not aimed at reproducing all the
\lx/\lb\sp variation, down to the lowest \lx/\lb\sp values observed (log
\lx/\lb$\sim -4$, with \lx\sp and \lb\sp in erg s$^{-1}$), but stop at log 
\lx/\lb$\sim -2.8$. Probably it is 
not possible to reproduce the lowest \lx/\lb\sp values by
simply truncating global inflows, because one continues to obtain galaxies quite
rich in hot gas. We note here also that NGC~3923 {\it is} a group-dominant
elliptical, but does not show  a very large hot gas content (log $L_{X}/
L_B\approx -3.3$).

The second way of explaining the scatter, through different dynamical phases
of the gas flows, regulated by {\it internal} agents, has the advantage of being a {\it
general} explanation, i.e., of applying to all the galaxies, regardless of
their environment (accretion is always needed for the extremely 
X-ray bright galaxies). At fixed \lb, any of the flow phases 
ranging from winds to subsonic outflows to partial and
global inflows, can be found at the present epoch, depending 
on the various depths and shapes of the potential well of the galaxies
 (Ciotti \etal 1991, Pellegrini \& Ciotti 1998). In
this way the large scatter in \lx\sp is easily accounted for: in the X-ray
bright galaxies the soft X-ray emitting gas dominates the emission,
being in the inflow phase, that resembles a cooling flow; in the X-ray
faint galaxies the hard stellar emission dominates, these being in the
wind phase; in intermediate \lx/\lb\sp galaxies, the hot gas is in the
outflow or partial wind phase, and the amount of soft emission
varies from being comparable to that of the stars, to being dominating.
In this scenario a crucial role is played by the SNIa explosions, that 
heat the flow in an extent sometimes large enough to drive all or part of the gas out of the
galaxies, and by the evolution of the explosion rate\footnote{The hot gas can
be expelled from the galaxies also when a central massive black hole is present,
because the gas flows are found to be unstable due to Compton heating
(Ciotti \& Ostriker 1997).}.
Numerical simulations using the updated rate given recently by Cappellaro et 
al. 1997, which is reduced with respect to that used
by Ciotti et al. (1991), 
 show that the partial wind phase is the most frequent, and that a large
scatter in the stagnation radius corresponds to a large scatter in the
amount of hot gas (Pellegrini \& Ciotti 1998). The problem with this scenario
is represented by the puzzle of the extremely low hot gas iron abundances
revealed by {\it ASCA} (Sect. 4.1.1).

\section { Conclusions}				

We have analysed the X-ray properties of the E4 galaxy NGC~3923 over the
energy range (0.5--10) keV, using {\it BeppoSAX} data.
Our findings are as follows.

 \begin{enumerate}
      \item The superposition of two thermal 
components of \ts$\approx 0.4$ keV, and \th$\approx 6-8$ keV, is the most 
reasonable representation of the spectral data. 
The heavy element abundances cannot be constrained due to the 
uncertainties in the spectral data; at the best fit they are very 
subsolar (except for the cooling flow model), and much lower than the 
stellar mean abundance.

\item The two components have roughly comparable fluxes in the
(0.5--4.5) keV band, while the hard component amounts to $\sim 3/5$ of the 
total (0.5--10) keV flux. 

\item The temperature of the softer component suggests as its origin the
emission of a hot gas, the origin
of the hard component is likely the integrated emission of LMXBs. In fact,
in addition to the spectral shape, also 
the amount of the hard emission is consistent with that predicted for
stellar sources in NGC~3923.

\item \ts\sp is close to the kinetic temperature of the stars in NGC~3923,
but some additional heating is needed.
Since \lx\sp is less than predicted by a steady state
cooling flow model, for a galaxy of an optical luminosity as high as that
of NGC~3923, it is suggested that a large fraction of the stellar mass loss
was removed by internal agents, such as the heating of SNIa's explosions,
and that this process was helped by the flat mass distribution of the 
galaxy. Another possibility, that has to be explored with numerical 
simulations, is that a substantial amount of hot gas was lost 
as a consequence of the episod of interaction or merger, with a much smaller 
galaxy, which gave origin to the system of shells visible in the optical. 

\item The origin of the X-ray emission in galaxies of
 low and medium \lx/\lb\sp is finally reviewed. The detailed study of these
galaxies is crucial to establish which factor plays the major role in
lowering the amount of hot gas, and so to explain the large scatter in the
\lx/\lb\sp plane. Is this factor to be linked to
external agents, as stripping by ambient gas or interactions with
other galaxies, or to internal heating mechanisms, such as SNIa explosions or 
accreting supermassive black holes? The galaxy discussed here is not surrounded
 by a dense medium, so that ram pressure stripping cannot be invoked;
but galaxy interactions clearly took place, and so cannot be excluded as causes of the loss of hot gas.
In general, since many other low and medium \lx/\lb\sp galaxies reside in small groups,
in which the density of the environment is presumably low or very low, it is
concluded that the most effective mechanism, in order to explain the
\lx/\lb\sp of these galaxies 
always with environmental effects, must be galaxy interactions.
But it is remarked that in general low or medium \lx/\lb\sp values are also 
shown by galaxies that reside in a region where
the galaxy density is not particularly high, and where also galaxies of
high \lx/\lb\sp are found.

\item Another result emerged here is that NGC~3923 is the dominant elliptical 
of its group, but has just a medium value of \lx/\lb. So,
an optically dominant galaxy does not always have
a high hot gas content.

\end{enumerate}

\par\noindent
NGC~3923 also belongs to a program of investigating the nature of
the X-ray emission in flat galaxies with very well known internal kinematics
and photometry (see Pellegrini et al. 1997), for which the mass profile can
be derived with accuracy.
Two-dimensional numerical simulations are in program to study in detail the
hot gas evolution in these galaxies, including the effect of galaxy 
interactions for this particular case.

\begin{acknowledgements}
This research has made use of SAXDAS linearized and cleaned event files 
(rev0) produced at the BeppoSAX Science Data Center. F. Fiore is warmly
thanked for enlightenments concerning SAX instrumental properties and data 
analysis.
Stimulating  discussions with F. Brighenti, L. Ciotti, A. D'Ercole and
G. Trinchieri are acknowledged. The referee, D. Buote, is also thanked for 
comments that improved the paper. 
\end{acknowledgements}

\end{document}